\documentclass[onecolumn,eqsecnum,showpacs,twoside,pra]{revtex4}
\usepackage{mathrsfs}
\usepackage{amssymb, amsbsy, amsmath, latexsym, dsfont, array, layout, graphicx,mathrsfs,color}

\newcommand{\ket}[1]{\left|{#1}\right\rangle}
\newcommand{\bra}[1]{\left\langle{#1}\right|}

\begin{document}

\title{Non-Markovian Dynamics of Spin Squeezing}
\author{Peng Xue}
\affiliation{Department of Physics, Southeast University, Nanjing
211189, China}
\date{\today}

\begin{abstract}
We evaluate the spin squeezing dynamics of $N$ independent
spin-$1/2$ particles with exchange symmetry. Each particle couples
to an individual and identical reservoir. We study the time
evolution of spin squeezing under the influence of different
decoherence. The spin squeezing property vanishes with evolution
time under Markovian decoherence, while it collapses quickly and
revives under non-Markovian decoherence. As spin squeezing can be
regarded as a witness of multipartite entanglement, our scheme shows
the collapses and revivals of multipartite entanglement under the
influence of non-Markovian decoherence.

\end{abstract}


\maketitle

\section{Introduction}
Quantum correlation has been playing a central role in quantum
information science and has also found many promising applications
such as achieving interferometric \cite{C81,Y86,KU91,UK01} and
enhancing the signal-to-noise ratio in spectroscopy
\cite{WBIMH92,AL02} beyond the standard quantum noise limit. The
spin squeezed state is one kind of quantum correlated states
\cite{CL09,X2012} with reduced fluctuations in one of the collective
spin components, which can be used to improve the precision of
atomic interferometers and atomic clocks. As an important quantum
correlation, entanglement is based on the superposition principle
combined with the Hilbert space structure, while spin squeezing is
originated from another fundamental principle of quantum
mechanics---the uncertainty principle. It has been proved that the
spin squeezing is closely related to and implies quantum
entanglement~\cite{KU91,WBIMH92,SDCZ01,WS03,TKGB07}. As a measure of
multipartite entanglement spin squeezing is relative easy to be
operated and measured.

To evaluate the potential application of quantum correlations such
as spin squeezing and entanglement, it is therefore essential to
include a realistic description of noise in experiments of
interests~\cite{BP02}. The dynamics of entanglement in open systems
has been broadly studied~\cite{XLG+10}. A peculiar aspect of the
entanglement dynamics is the well known ``entanglement sudden death"
phenomenon~\cite{ESD1,ESD2,ESD3} and recently the ``sudden death" of
spin squeezing during a Markovian process has been
investigated~\cite{WMLPN10,YMWN12}. The unidirectional flow of
information in which the decoherence and noise act consistently,
characterizes a Markovian process. However, there are some systems
such as condensed-matter systems which are strongly coupled to the
environment and the coupling leads to a different regime where
information also flows back into the system from the surroundings,
which characterizes a non-Markovian process. Memory effects caused
by the information flowing back to the system during a non-Markovian
process can temporarily interrupt the monotonic increases or
decreases of distinguishability such as spin squeezing parameter. In
this paper we study the spin squeezing dynamics of $N$ independent
spin-$1/2$ particles with exchange symmetry which are coupled to
individual and identical non-Markovian decoherence channels and show
the collapses and revivals of spin squeezing.

\section{Spin Squeezing Definitions}

We consider an ensemble of $N$ two-level particles with lower
(upper) state $\ket{\downarrow}$ ($\ket{\uparrow}$). Adopting the
nomenclature of spin-$1/2$ particles, we introduce the total angular
momentum
\begin{equation}
\vec{J}=\sum_{j=1}^N\vec{S}_j,
\end{equation}
where $\vec{S}_z^j=\frac{1}{2}\hat{\sigma}_z^j=
\frac{1}{2}\left(\ket{\uparrow}_j\bra{\uparrow}-\ket{\downarrow}_j\bra{\downarrow}\right)$.
At this point, it is convenient to introduce the following
definition of spin squeezing parameter \cite{WBIMH92,AP94}
\begin{equation}
\xi^2=\frac{N \big(\Delta
J_{\vec{n}_\bot}\big)_{min}^2}{\langle\vec{J}\rangle^2}.
\end{equation}
Here the minimization is over all directions denoted by
$\vec{n}_\bot$, perpendicular to the mean spin direction
$\vec{n}=\langle\vec{J}\rangle/|\langle\vec{J}\rangle|$. If
$\xi^2<1$ is satisfied, the spin squeezing occurs and the $N$-qubit
state is entangled.

There are also other definitions for spin squeezing parameters which
might show different sensitivities to the decoherence channels. We
introduce another parameter defined by T\'{o}th et al.~\cite{TKGB07}
\begin{equation}
\xi'^2=\frac{\lambda_{min}}{\langle\vec{J}^2\rangle-N/2},
\end{equation}
where $\lambda_{min}$ is minimum eigenvalue of the matrix
$\Gamma=(N-1)\Upsilon+C$ with $\Upsilon_{kl}=C_{kl}-\langle
J_k\rangle \langle J_l\rangle$ for $k,l\in\{x,y,z\}$ the covariance
matrix and $C=\left[C_{kl}\right]$ with $C_{kl}=\langle J_l
J_k+J_kJ_l\rangle/2$ is the global correlation matrix.

\section{One-Axis Twisted Spin Squeezed States}

Now we introduce one kind of spin squeezed states---one-axis twisted
spin squeezed states. Consider an ensemble of $N$ spin-$1/2$
particles with exchange symmetry and assume that the dynamical
properties of the system can be described by collective operators
$J_\alpha$, $\alpha=x,y,z$. The one-axis twisting
Hamiltonian~\cite{KU93,S89,LXJY11} is an Ising-type Hamiltonian
\begin{equation}
\hat{H}=\sum_{j\neq
k}\frac{1}{4}f(j,k)\left(\mathbb{I}-\hat{\sigma}_z^j\right)
\otimes\left(\mathbb{I}-\hat{\sigma}_z^k\right),
\end{equation}
which involves all pairwise interactions with coupling constant
$f(j,k)$.

The one-axis twisted spin squeezed state \cite{WM02,TIT+05,JLL09}
can be prepared by the evolution of the above Hamiltonian
\begin{eqnarray}
\ket{\psi_t}=\exp\left(-{i}\hat{H}t\right)\ket{+}^{\otimes
N}=\prod_{j\neq
k}\exp\left[-\frac{{i}}{4}f(j,k)t\hat{\sigma}_z^j\hat{\sigma}_z^k\right]\ket{+}^{\otimes
N},
\end{eqnarray}
where
$\ket{+}=\left(\ket{\uparrow}+\ket{\downarrow}\right)/\sqrt{2}$. If
we choose the evolution time to satisfy $f(j,k)t=m\pi$ with $m$ an
integer, the state $\ket{\psi_t}$ is a product state. If
$f(j,k)t=\left(2m+1\right)\pi/2$, $\ket{\psi_t}$ becomes a graph
state. For $0<f(j,k)t<\pi/2$, $\ket{\psi_t}$ is a one-axis twisted
spin squeezed state characterized by spin squeezing parameter $\xi$
($\xi'$).

The spin squeezing parameter $\xi$ of the one-axis twisted spin
squeezed state with all coupling coefficients satisfying
$f(j,k)t=\alpha$ takes this form
\begin{equation}
\xi^2=\frac{1-\left(N-1\right)\left[\sqrt{A^2+B^2}-A\right]/4}{\cos^{2N-2}\alpha},
\end{equation}
where
\begin{equation}
A=1-\cos^{N-2}\left(2\alpha\right), B=4\sin\alpha\cos^{N-2}\alpha.
\end{equation}
The mean spin direction for the one-axis twisted spin squeezed state
is
\begin{equation}
\vec{n}=\left(\cos\left(N\alpha\right),\sin\left(-N\alpha\right),0\right),
\end{equation}
and the orthogonal direction is
\begin{equation}
\vec{n}_{\perp}=\left(-\cos\phi\sin\left(-N\alpha\right),\cos\phi\cos\left(N\alpha\right)
,\sin\phi\right).
\end{equation}
The minimum spin squeezing parameter with respect to $\alpha$ is
obtained $\xi\propto 1/N^{1/3}$ shown in Fig.~1.

The spin squeezing parameter with another definition $\xi'$ of the
one-axis twisted spin squeezed state above takes the form
\begin{equation}
\xi'^2=\frac{\text{min}(a,b)}{(1-1/N)(1+\cos^{N-2}2\alpha)/2+1/N},
\end{equation}
where \begin{align}&a=1-(N-1)(\sqrt{A^2+B^2}-A)/4,\\
&b=1+(N-1)\left[(1+\cos^{N-2}2\alpha)/2-\cos^{2N-2}\alpha\right].\nonumber\end{align}

\section{Evolution of Spin Squeezing in The Presence of Decoherence}

For a single qubit coupled to a noisy channel which is described by
a thermal reservoir, the evolution of this qubit is governed by a
general master equation of a Lindblad form
\begin{equation}
\frac{d}{d t}\chi={i}\left[\hat{H}_r,\chi\right]+\mathcal{L}\chi,
\end{equation}
where the reference system is
\begin{equation}
\hat{H}_{r}=\frac{\Delta}{2}\sum_{j=1}^N\hat{\sigma}_z^j
\end{equation} with $\Delta$ is the strength of the external field.
The optimal spin squeezed states are eigenstates of the
Hamiltonian~\cite{R03}. Whereas, the incoherent processes are
described by the superoperator $\mathcal{L}$:
\begin{eqnarray}
\label{l}
\mathcal{L}\chi=&&-\frac{b}{2}(1-s)\left[\hat{\sigma}_+\hat{\sigma}_-\chi+
\chi\hat{\sigma}_+\hat{\sigma}_--2\hat{\sigma}_-\chi\hat{\sigma}_+\right]\\
&&-\frac{b}{2}s\left[\hat{\sigma}_-\hat{\sigma}_+\chi+\chi\hat{\sigma}_-\hat{\sigma}_+
-2\hat{\sigma}_+\chi\hat{\sigma}_-\right]\nonumber\\
&&-\frac{2c-b}{8}
\left[2\chi-2\hat{\sigma}_z\chi\hat{\sigma}_z\right],\nonumber
\end{eqnarray}
with $\hat{\sigma}_\pm=\left(\hat{\sigma}_x\pm
{i}\hat{\sigma}_y\right)/2$. For $b=0$, $c=\gamma$ and an arbitrary
$s$, the generator Eq.~(\ref{l}) describes the coupling between the
qubit and a dephasing channel. For $s=1/2$ and $b=c=\gamma$, the
qubit is coupled to a depolarizing channel. Whereas, for $s=1$ and
$b=2c=\gamma$, that is coupled to a decay channel (pure damping
channel).

Equivalently, one can use the resulting completely positive map
$\mathcal{E}$ with $\chi'=\mathcal{E}\chi$ to describe the
decoherence channels
\begin{equation}
\mathcal{E}\chi=\sum_{j=0}^3 p_j\hat{\sigma}_j\chi\hat{\sigma}_j.
\end{equation}
with $\chi$ a density matrix for a single-qubit state and
$\sum_{j=0}^3p_j=1$. These decoherence channels are of practical
interests in quantum information science. This class contains for
example: (i) for $p_0=\left(1+3\kappa^2\right)/4$ and
$p_1=p_2=p_3=\left(1-\kappa^2\right)/4$ with $\kappa=e^{-\gamma t}$
$\mathcal{E}$ describing a depolarizing channel; (ii) for
$p_0=\left(1+\kappa^2\right)/2$, $p_1=p_2=0$ and
$p_3=\left(1-\kappa^2\right)/2$ a dephasing channel. Finally, the
decay channel is described
\begin{equation}
\mathcal{E}\chi=E_0\chi E_0^\dagger+E_1\chi E_2^\dagger,
\end{equation}
with the Kraus operators $E_0=\left(
                                \begin{array}{cc}
                                  1 & 0 \\
                                  0 & \kappa \\
                                \end{array}
                              \right)$ and $E_1=\left(
                                \begin{array}{cc}
                                  0 & \sqrt{1-\kappa^2} \\
                                  0 & 0 \\
                                \end{array}
                              \right)$~\cite{HDB04}.

The quantum master equations with the time-local structures are also
very useful for the description of non-Markovian processes. Suppose
we have a time-local master equation of the form
\begin{equation}
\frac{d}{dt}\chi={i}\left[\hat{H}_r,\chi\right]+\mathcal{K}(t)\chi,
\end{equation}
where $\mathcal{K}(t)$ is a time-dependent generator and takes the
similar form in Eq.~(\ref{l}) with time-dependent parameter
$\gamma(t)$.

If the relaxation rate $\gamma(t)$ is positive, the generator
$\mathcal{K}(t)$ takes the Lindblad form for each fixed $t\geq 0$.
In the Markovian regime, information encoded in the qubit state
leaks to its surroundings and $|\kappa(t)|$ is a monotonically
decreasing function of times. In the non-Markovian environment, in
contrast, information also flows back into the system of qubit state
and a revival of distinguishability (here the spin squeezing
parameter) can be observed in the time evolution. With the
definition of non-Markovianity~\cite{BLP09}, we see that an increase
of $|\kappa(t)|$ leads to a negative rate $\gamma(t)$ in the
generator $\mathcal{K}(t)$. For example, we consider the case of a
Lorentzian reservoir spectral density which is on the resonance with
the spin qubit transition frequency and leads to an exponential two
point correlation function
\begin{equation}
f(t)=\frac{1}{2}\eta_0\gamma e^{-\gamma t}.
\end{equation}
The rate $\kappa(t)$ is defined as the solution of the
integrodifferential equation
\begin{equation}
\frac{{d}}{{d}t}\kappa(t)=-\int_{0}^{t}{d}t'f(t-t')\kappa(t')
\end{equation} corresponding to an initial condition $\kappa(0)=1$. The
parameter $\gamma$, defining the spectral width of the coupling
between the qubit and reservoir, is connected to the reservoir
correlation time $\tau\approx \gamma^{-1}$. For a weak coupling
regime $\eta_0<\gamma/2$, the relaxation time is greater than the
reservoir correlation time and the behavior of $\kappa(t)$ is a
Markovian exponential decay. In the strong coupling regime
$\eta_0>\gamma/2$, the reservoir correlation time is greater than
the relaxation time and non-Markovian effects become
relevant~\cite{XLG+10,BLP09,WECC08,RHP10,LPB10,RBH97,M+05,XLZ+10,L+11,T+12}.
Thus we obtain
\begin{equation}
\kappa(t)=e^{\frac{-\gamma t}{2}}\left[\cos\left(\frac{dt}{2}\right)
+\frac{\gamma}{d}\sin\left(\frac{dt}{2}\right)\right],
\end{equation}
where $d=\sqrt{2\eta_0\gamma-\gamma^2}$.

We would be interested in the effect of decoherence on the spin
squeezing properties of a system including $N$ two-level particles.
A decoherence channel individually coupling to each qubit is
considered in this paper, and the evolution of the $k$th qubit is
described by the map $\mathcal{E}_k$ with Pauli operators
$\hat{\sigma}_j$ ($j=0,1,2,3$) acting on qubit $k$. We are
interested in the dynamical evolution of a given one-axis twisted
spin squeezed state $\psi$ of $N$ qubits in the presence of
decoherence. The initial state $\psi$ evolves to a mixed state
$\rho(t)$ given by
\begin{equation}
\rho(t)=\mathcal{E}_1\mathcal{E}_2...\mathcal{E}_N\ket{\psi_t}\bra{\psi_t}.
\end{equation}

For a one-axis twisted spin squeezed state, we consider three kinds
of decoherence channels. The modified mean spin direction and the
orthogonal direction are calculated as
\begin{eqnarray}
&&\vec{n}'=\left(\cos\left(\Delta t-N\alpha\right),\sin\left(\Delta
t-N\alpha\right),0\right),\\
&&\vec{n}'_\perp=\left(-\cos\phi\sin\left(\Delta
t-N\alpha\right),\cos\phi\cos\left(\Delta
t-N\alpha\right),\sin\phi\right).\nonumber
\end{eqnarray}
Under the influence of the individual dephasing channels which are
the main type of decoherence for a spin ensemble, the spin squeezing
parameter of the one-axis twisted spin squeezed state evolves to
\begin{equation}
\xi^2_{deph}(t)=\frac{\zeta}{\cos^{2N-2}\alpha},
\end{equation} where
\begin{eqnarray}
\zeta=1+\frac{1}{4}\kappa^2\left(t\right)\left(N-1\right)\left(A-\frac{A^2}{\sqrt{A^2+B^2}}\right)
-\frac{1}{4}\kappa\left(t\right)\left(N-1\right)\frac{B^2}{\sqrt{A^2+B^2}}
\end{eqnarray}
Then the spin squeezing parameter of the one-axis twisted spin
squeezed state evolves to
\begin{equation}
\xi^2_{depol}(t)=\frac{\zeta}{\kappa^2\left(t\right)\cos^{2N-2}\alpha}.
\end{equation}
under depolarizing. Whereas, the spin squeezing parameter of the
one-axis twisted spin squeezed state evolves to
\begin{equation}
\xi^2_{damp}(t)=\frac{\zeta}{\Big\{
\kappa\left(t\right)\cos^{N-1}\alpha+\left[1-\kappa\left(t\right)\right]\Big\}^2}.
\end{equation} under damping.

With the other definition shown in Eqs. (2.3) and (3.7), the spin
squeezing parameter $\xi'$ of the one-axis twisted spin squeezed
states coupled to individual dephasing, depolarizing and damping
channels evolves as followings
\begin{align}
&\xi'^2_{deph}(t)=\frac{\zeta}{(1-1/N)\{\kappa(t)^2+\left[1-\kappa(t)^2\right](1+\cos^{N-2}2\alpha)/2\}+1/N},\\
&\xi'^2_{depol}(t)=\frac{\zeta}{(1-1/N)\kappa(t)^2+1/N}, \\
&\xi'^2_{damp}(t)=\frac{\zeta}{1+(1-1/N)\kappa(t)(1-\kappa(t))\left[1-\cos^{N-1}\alpha+(1+\cos^{N-2}2\alpha)/2\right]}.
\end{align}

The time evolution of spin squeezing parameters $\xi$ and $\xi'$ of
$10$ particles prepared initially in one-axis twisted spin squeezed
state coupled to individual dephasing, depolarizing and damping
channels are shown in Figs. 2-7, respectively. We compare the
evolution of spin squeezing under Markovian and non-Markovian
decoherence. Here we consider a Lorentzian reservoir and thus the
deocherence function $\kappa(t)$ can be written as exponential decay
$e^{-\gamma t/2}$ modified by a periodical time-dependent function
$\cos(dt/2)+\gamma/d\sin(dt/2)$. In short time regime, the spin
squeezing property collapses and revives under the influence of
either of three kinds of non-Markovian decoherence. With time
increasing, the part of the exponential decay becomes more important
and the spin squeezing property vanishes finally as that under
Markovian decoherence does. The spin squeezing evolving under
Markovian decoherence gives a lower bound of the envelope of that
evolving under non-Markovian decoherence shown in Figs 2-7. The two
parameters show different sensitivities to different decoherence
channels. Both parameters are robust to damping channel in Figs. 4
and 7 and the squeezing property represented by the parameters $\xi$
and $\xi'$ lasts for $t=1000$. Coupled to the other two decoherence
channels (dephasing and depolarizing channels), the parameter $\xi$
is more sensitive than $\xi'$. In Figs. 2 and 5, the disappearance
times for spin squeezing representing by the parameters $xi$ and
$\xi'$ under dephasing are about $t=259.99$ and $t=318.13$,
respectively. In Figs. 3 and 6, the disappearance times for spin
squeezing representing by $\xi$ and $\xi'$ under depolarizing are
about $t=68.76$ and $t=106.01$, respectively. Both parameters $\xi$
and $\xi'$ are more sensitive to the depolarizing.

\section{Conclusion}
In summary, we study the dynamics processes of the spin squeezing of
a spin ensemble in which each spin is coupled to an independent and
identical decoherence channel. We analytically calculate the
dynamics of the spin squeezing parameters under three different
types of decherence. As we know the Heisenberg scaling $1/N$ in the
decoherence-free case can be achieved. In the presence of Markovian
decoherence the spin squeezing property of one-axis twisted states
vanishes with the evolution time. Whereas, in the presence of
non-Markovian decohernce and in the short time limit, the spin
squeezing property collapses and revives with the evolution time due
to short-time memory effect during non-Markovian processing. With
time increasing, the spin squeezing vanishes finally even under
non-Markovian decoherence. As spin squeezing can be regarded as a
measure/witness of multiqubit entanglement, thus our scheme for the
first time shows the collapses and revivals of multiqubit
entanglement under non-Markovian decoherence.

\begin{acknowledgments}
We would like to thank Yongsheng Zhang and Xiangfa Zhou for useful
conversations. This work has been supported by the National Natural
Science Foundation of China under Grant Nos 11004029 and 11174052,
the Natural Science Foundation of Jiangsu Province under Grant No
BK2010422, the Ph.D.\ Program of the Ministry of Education of China,
the Excellent Young Teachers Program of Southeast University and the
National Basic Research Development Program of China (973 Program)
under Grant No 2011CB921203.
\end{acknowledgments}

\newpage
\begin{figure}
   \includegraphics[width=.45\textwidth]{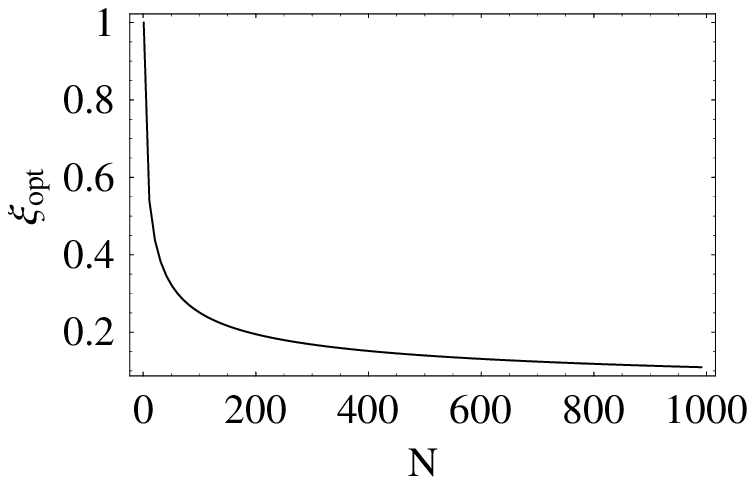}
   \caption{The plot of the spin squeezing parameter $\xi$ for a one-axis twisted
   spin squeezed state v.s. the number of the qubits $N$ optimized
   with respect to $\alpha$.}
   \label{figure3}
\end{figure}

\begin{figure}
   \includegraphics[width=.45\textwidth]{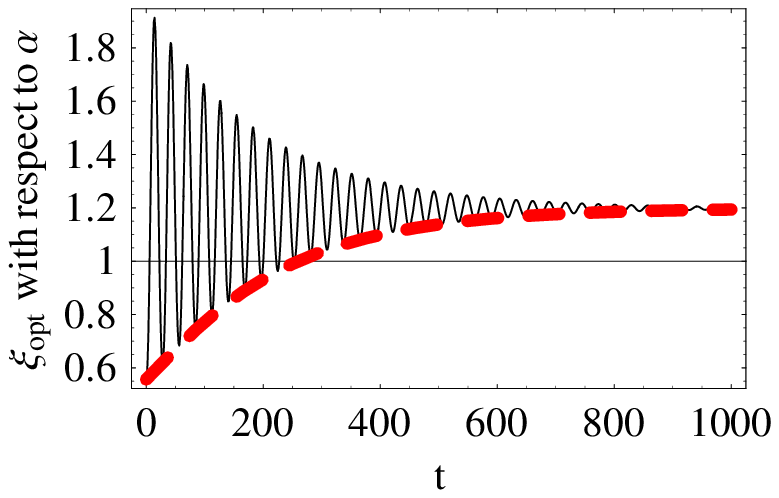}
   \caption{The time evolution of spin squeezing of a one-axis twisted
   spin squeezed state coupled to non-Markovian dephasing channel representing by $\xi$ v.s. the time $t$ with $N=10$,
   $\gamma=0.01$ and $\eta_0=10$ (in black solid line).
   For comparison, we plot the evolution of the spin squeezing under Markovian dephasing with $\kappa(t)=e^{-0.005t}$ (in red dashed line).}
   \label{dephasing}
\end{figure}

\begin{figure}
   \includegraphics[width=.45\textwidth]{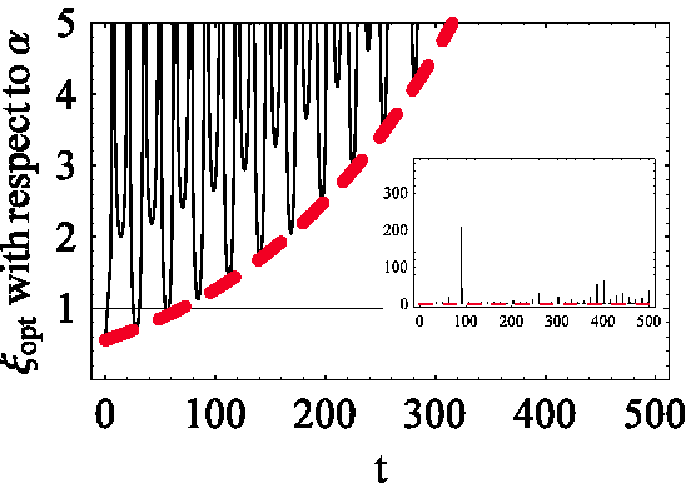}
   \caption{The time evolution of spin squeezing of a one-axis twisted
   spin squeezed state coupled to non-Markovian depolarizing channel representing by $\xi$ v.s. the time $t$ with $N=10$,
   $\gamma=0.01$ and $\eta_0=10$ (in black solid line).
   For comparison, we plot the evolution of the spin squeezing under Markovian depolarizing with $\kappa(t)=e^{-0.005t}$ (in red dashed line).}
   \label{depolarizing}
\end{figure}

\begin{figure}
   \includegraphics[width=.45\textwidth]{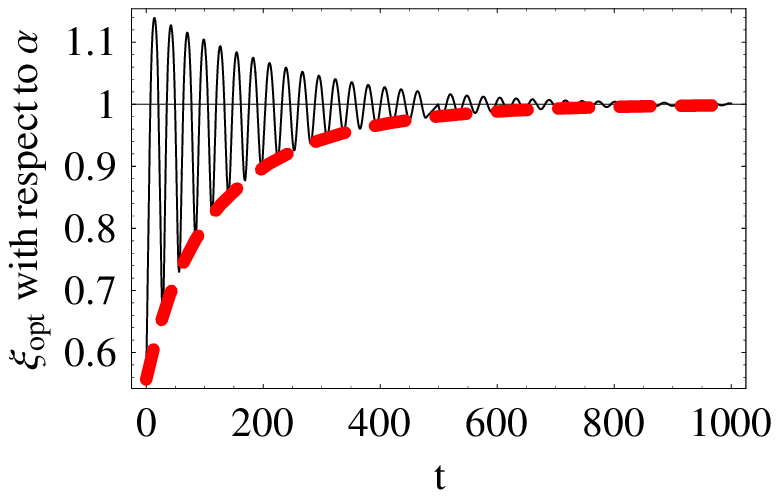}
   \caption{The time evolution of spin squeezing of a one-axis twisted
   spin squeezed state coupled to non-Markovian damping channel representing by $\xi$ v.s. the time $t$ with $N=10$,
   $\gamma=0.01$ and $\eta_0=10$ (in black solid line).
   For comparison, we plot the evolution of the spin squeezing under Markovian damping with $\kappa(t)=e^{-0.005t}$ (in red dashed line).}
   \label{damping}
\end{figure}

\begin{figure}
   \includegraphics[width=.45\textwidth]{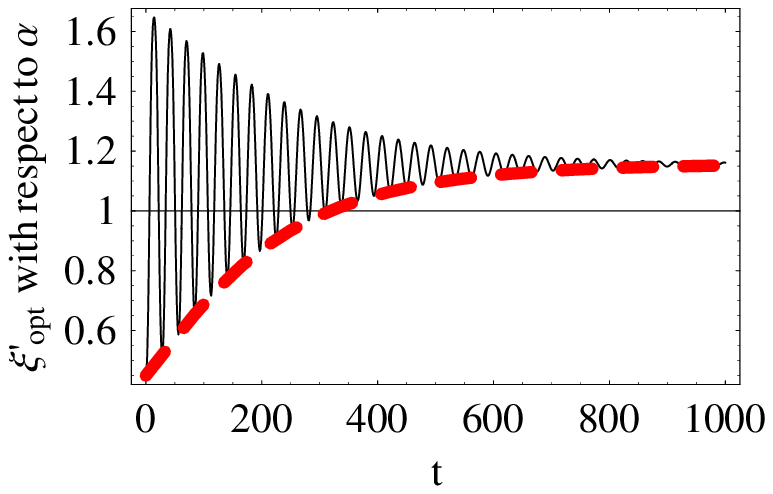}
   \caption{The time evolution of spin squeezing of a one-axis twisted
   spin squeezed state coupled to non-Markovian dephasing channel representing by $\xi'$ v.s. the time $t$ with $N=10$,
   $\gamma=0.01$ and $\eta_0=10$ (in black solid line).
   For comparison, we plot the evolution of the spin squeezing under Markovian dephasing with $\kappa(t)=e^{-0.005t}$ (in red dashed line).}
   \label{dephasing}
\end{figure}

\begin{figure}
   \includegraphics[width=.45\textwidth]{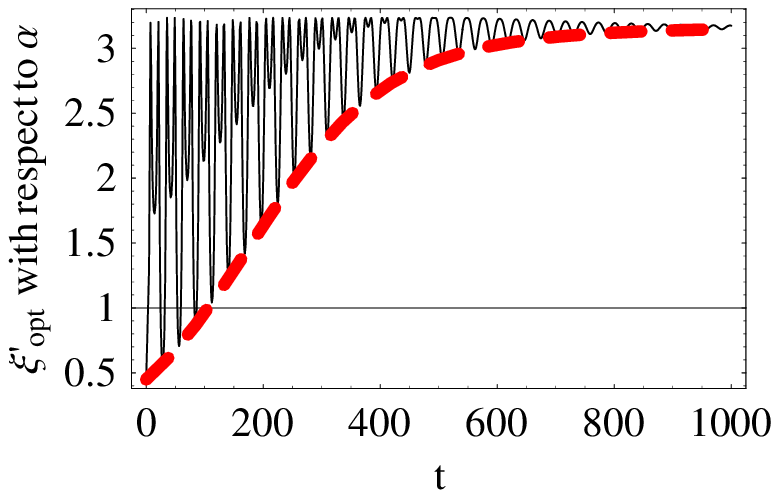}
   \caption{The time evolution of spin squeezing of a one-axis twisted
   spin squeezed state coupled to non-Markovian depolarizing channel representing by $\xi'$ v.s. the time $t$ with $N=10$,
   $\gamma=0.01$ and $\eta_0=10$ (in black solid line).
   For comparison, we plot the evolution of the spin squeezing under Markovian depolarizing with $\kappa(t)=e^{-0.005t}$ (in red dashed line).}
   \label{depolarizing}
\end{figure}

\begin{figure}
   \includegraphics[width=.45\textwidth]{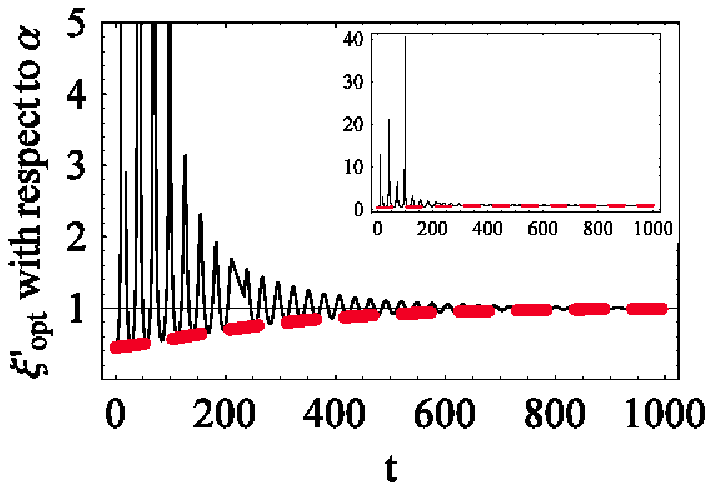}
   \caption{The time evolution of spin squeezing of a one-axis twisted
   spin squeezed state coupled to non-Markovian damping channel representing by $\xi'$ v.s. the time $t$ with $N=10$,
   $\gamma=0.01$ and $\eta_0=10$ (in black solid line).
   For comparison, we plot the evolution of the spin squeezing under Markovian damping with $\kappa(t)=e^{-0.005t}$ (in red dashed line).}
   \label{damping}
\end{figure}

\end{document}